\documentclass{emulateapj}

\usepackage{color,colortbl}
\usepackage{multirow}
\usepackage{natbib}
\usepackage{graphicx}
\usepackage{epstopdf}
\usepackage{amssymb}
\usepackage{amsmath}
\usepackage{gensymb}
\usepackage{diagbox}

\newcommand\T{\rule{0pt}{2.6ex}}       % Top strut
\shorttitle{Dynamical Evolution of Alpha Centauri}
\shortauthors{Worth \& Sigurdsson}

\begin{document}

\title{Effects of Proxima Centauri on Planet Formation in Alpha Centauri}

\author{R. Worth\altaffilmark{1,2} and S. Sigurdsson\altaffilmark{1,2}}

\affil{The Astronomy \& Astrophysics Department, \\
	Pennsylvania State University, University Park, PA 16802}

\altaffiltext{1}{Penn State Astrobiology Research Center, Penn State.}
\altaffiltext{2}{Center for Exoplanets and Habitable Worlds, Penn State.}

%%%%%%%%%%%%%%%%%%%%%%%%%%%%%%%%%%%%%%%%%%%%%%%%%%%%%%%%%%%%%%%%%%%%%%%%%%%%%%%

\begin{abstract}
Proxima Centauri is an M dwarf approximately 15,000 AU from the Alpha Centauri binary, comoving and likely in a loosely bound orbit. Dynamic simulations show this configuration can form from a more tightly bound triple system. As our nearest neighbors, these stars command great interest as potential planet hosts, and the dynamics of the stars govern the formation of any planets within the system. Here we present a scenario for the evolution of Alpha Centauri A and B and Proxima Centauri as a triple system. Based on N-body simulations, we determine this pathway to formation is plausible, and we quantify the implications for planet formation in the Alpha Centauri binary. We expect this formation scenario may have truncated the circumstellar disk slightly more than a system that formed in the current configuration, but that it most likely does not prevent terrestrial planet formation. We simulate planet formation in this system and find that in most scenarios, two or more terrestrial planets can be expected around either Alpha Centauri A or B, orbiting in a region out to approximately 2 AU, assuming planetesimals and planetary embryos are able to first form in the system. Additionally, terrestrial planet formation and stability in Proxima Centauri's habitable zone is also plausible. However, an absence of planets around these stars may be indicative of highly disruptive stellar dynamics in the past. 
\end{abstract}

%%%%%%%%%%%%%%%%%%%%%%%%%%%%%%%%%%%%%%%%%%%%%%%%%%%%%%%%%%%%%%%%%%%%%%%%%%%%%%%
%%%%%%%%%%%%%%%%%%%%%%%%%%%%%%%%%%%%%%%%%%%%%%%%%%%%%%%%%%%%%%%%%%%%%%%%%%%%%%%
\section{Introduction}

Our nearest neighbor, the M-dwarf Proxima Centauri, is thought to be tenuously bound to the Alpha Centauri binary, forming an extremely wide triple system; although measurements are not precise enough to constrain the orbit, its proximity to Alpha Centauri along with its comoving velocity would be very unlikely in a passing, unconnected star \citep{2006AJ....132.1995W}. A detailed study of triple system dynamics \citep{2012Natur.492..221R} hypothesized that the three stars could have formed closer together, as part of a single system, and that dynamical interactions between them could then have led to Proxima's near-ejection onto its current highly eccentric path. 

There has also been significant interest in the possibility of planets in the Alpha Centauri system. As our nearest neighbors, these stars represent the best candidates for in-depth study, as well as the most likely target for a search for biomarkers or any distant future interstellar contact if habitable planets were to be found there. A hot Earth-sized planet has been reported around Alpha Centauri B \citep{2012Natur.491..207D}, although the detection has been disputed \citep{2013ApJ...770..133H, 2016MNRAS.456L...6R}. Other observational studies have ruled out the possibility of giant planets in the system \citep{2015MNRAS.450.2043D, 2001A&A...374..675E} and put constraints on the detectability of planets within the system \citep{2015IJAsB..14..305E, 2013ApJ...764..130E, 2008ApJ...679.1582G}. 

The gravitational forces of multiple stars introduce additional complications for planetary formation; however, the prevalence of stellar multiplicity means it is critical to understand such systems if we are to understand planet formation in the Universe as a whole \citep{2014arXiv1406.1357T}. Although planet searches initially avoided binaries due to their additional complications as observational targets as well as the reduced likelihood of planet formation in their more turbulent dynamical environments \citep{2010ASSL..366...19E}, many planets have now been discovered in binary systems, including a multiplanet system in the binary 55 Cancri \citep{2008ApJ...675..790F}. In some cases, planet searches discovered both a planet and a previously unknown companion star \citep{2009A&A...494..373M}. A planet has also been detected in 16 Cygni, a triple system consisting of two Sun-like stars and a red dwarf, similar to Alpha Centauri except that here the smaller star is part of the inner binary \citep{1997ApJ...483..457C}.

In addition to the multiple systems we see today, it is likely that even more stars were members of such systems when they formed. There is evidence to suggest many, if not most, stars form in bound multistellar systems which then eject members until they reach stable configurations, resulting mostly in singles and binaries but occasionally in higher-multiplicity systems \citep{2010ApJ...725L..56R, 2007prpl.conf..133G, 2014prpl.conf..267R}. This is consistent with the lower multiplicity rates of smaller stars, as they are more easily ejected than massive stars. The consequences of ``fly-by'' interactions between stars have been studied \citep{2015MNRAS.448..344L}, but multiple bound stars have different outcomes, as they repeatedly interact over multiple orbits. The protoplanetary disks around these stars, from which planets will eventually form, may be truncated or disturbed during these stellar interactions \citep{2007ApJ...660..807Q}.

Several groups have made theoretical studies of the disk or planet stability \citep{2015ApJ...799..147J, 2015ApJ...798...70R, 2015ApJ...798...69R, 2012A&A...539A..18M, 2012AstL...38..581P, 2009MNRAS.400.1936P, 1994ApJ...421..651A} and formation environment \citep{2010ApJ...708.1566X, 2009MNRAS.393L..21T, 2008MNRAS.388.1528T, 2008ApJ...679.1582G, 2002A&A...396..219B, 2002ApJ...576..982Q, 1997AJ....113.1445W} in Alpha Centauri. However, most of these studies assume the stars were in their current orbits. If Alpha Centauri exchanged energy with Proxima Centauri to allow the latter to reach its current orbit, the Alpha Centauri binary would have lost energy in the process, altering its own orbit as well. These interactions typically take place soon after the stars' formation, as does the formation of any circumstellar disks and protoplanets within them. Therefore, such interactions could have significant consequences for our assumptions about planet formation in this system. In \cite{2009MNRAS.393L..21T}, the possibility of wider initial orbits was examined in the context of formation in a cluster, finding that initially wider orbits could improve the ease of formation of planetesimals.

In this work, we seek to characterize the limits of the dynamical history of the Alpha Centauri star system, with particular interest in how it may affect any planet formation within that system. We will assume Proxima Centauri is on an eccentric, bound orbit, and that its current position at around 15,000 AU from the Alpha Centauri binary is likely in the long, slow, portion of its orbit relatively near apocenter. We constructed a large population of triple systems that could evolve into the current arrangement and simulate their interactions, looking for examples of Proxima (a.k.a. Alpha Centauri C) ending up on a wide, highly eccentric orbit. We examine two limiting cases: all stars at their present masses, so that A and B significantly outweigh C, thus minimizing the effects any energy exchanges have on the binary; and an equal-mass case which assumes that all three stars initially grew at similar rates, but that C was ejected when mass was still accreting and all three stars were its size (0.123 M$_\odot$), maximizing the effects on the binary system. This puts bounds on the amount by which the binary orbit could possibly change. Ultimately, we seek to understand whether planets may be present in Alpha Centauri, at what locations, and with what mass. The evolution of Proxima Centauri can affect this, but the effect could range from trivial to quite substantial depending on timing, so we bracket the range of possibilities to consider the range of outcomes.

%%%%%%%%%%%%%%%%%%%%%%%%%%%%%%%%%%%%%%%%%%%%%%%%%%%%%%%%%%%%%%%%%%%%%%%%%%%%%%%
\section{Simulations}

%------------------------------------------------------------------------------
\subsection{Numerical Method}
We performed suites of simulations in three different regimes to explore different stages of the system's evolution: three-body simulations of just the stars, simulations of two or three stars with a disk of test particles, and planetary formation simulations within a stellar binary system. These regimes are laid out in Table~\ref{simcases} for clarity. For all three, we used the hybrid symplectic N-body code {\sc mercury} \citep{1999MNRAS.304..793C}, which was designed for use on systems with a single, dominant central mass object. In some cases, a binary system will break the assumptions underlying the integration scheme and cause errors. By design, the first two regimes explored avoid these issues, in the first case due to the well-separated hierarchical nature of the systems, and in the second by the use of test particles. In order to correctly simulate planetary formation in a binary, however, we used a version of the hybrid symplectic integrator that has been modified for wide binary systems, as described in \cite{2002AJ....123.2884C} and used in \cite{2002ApJ...576..982Q}.

%- - - - - - - - - - - - - - - - - - - - - - - - - - - - - - - - - - - - - - - 
\begin{table*}[tbp]
\begin{center}
\caption{Simulation Regimes Used \label{simcases} }
\begin{tabular}{l|llll}
  \hline
 Regime & Simulations & Algorithm & Cases  & Features \T \\
  \hline
Stars   & 1522        & hybrid    & Early  & Three stars (all 0.123 M$_{\odot)}$) \T \\
        &             &           & Late   & Three stars (1.103, 0.934, and 0.123 M$_{\odot}$) \\
Disk    & 224         & hybrid    & Binary            & Stars A and B plus disk particles, \T \\
        &             &           & Triple            & Stars A, B, and C plus disk particles \\
Planets & 36          & wb        & various parameter & $\sigma_0$ = \{2.8, 3.1\} AU, \T \\
        &             &           & combinations      & $r_{tr}$ = \{0.3, 1, 3\} \\
\end{tabular}
\tablecomments{The different sets of simulations presented described in this paper, the number of simulations in each set, and the cases considered within each. All simulations were performed using a hybrid-symplectic algorithm in {\sc mercury}, but the planet formation simulations used a modified version of the algorithm intended for wide-binary systems.}
\end{center} 
\end{table*}
%- - - - - - - - - - - - - - - - - - - - - - - - - - - - - - - - - - - - - - - 

Our first regime follows the methods of \cite{2012Natur.492..221R} to examine the three-body interactions of Alpha Centauri A and B and Proxima Centauri in 1,522 randomized simulations. We added physics to {\sc mercury}'s user module that includes the gravitational potential of the nascent molecular cloud core from which the stars form, as in \cite{2012Natur.492..221R}. We omit accretion of gas directly onto the stars, but otherwise use parameters similar to theirs: 10 M$_{\odot}$ cloud core mass with a Plummer sphere potential of $M/(r^2+R^2)^{1/2}$, where $M$ is the cloud mass in M$_{\odot}$, $R$ is the cloud core radius, 7,500 AU, and $r$ is the distance of the object in question from the cloud center. The cloud mass disperses linearly over 440,000 years. In addition, because of the very large spatial separations involved, we included Galactic tidal forces using code provided by Dimitri Veras, as described in \cite{2013MNRAS.430..403V}. The modification described in \cite{2013AsBio..13.1155W}, which improves the criteria for determining when an object impacts the central body, was also included but is generally not significant for simulations which do not involve satellites.

Simulations were initially run for $10^3$ years, then checked for ejections. If none were found, the stop time was increased by an order of magnitude and the simulation was resumed. This process was repeated up to a simulation length of $10^{9}$ years. Systems surviving the full length were then counted as ``surviving'' if the eccentricities of each object were less than 1.0, i.e. the stars were on bound ellipsoidal orbits. The simulation does not count stars as ``ejected'' until they go beyond some outer bound, which was set to 100,000 AU to allow for the large orbits desired, so it is possible for objects to be on ejection orbits but not yet counted as ``ejected.'' It was also possible for stars to have orbits with apocenters greater than 100,000 AU, in which case they were falsely counted as ejected while still stable; however, at this scale we expect the orbit would be quickly destabilized by Galactic tides if it were integrated further. 

Also note, the orbital parameters output by {\sc mercury} assume all orbits are with respect to the designated ``central'' object, which is not the case here. Semimajor axis, eccentricity, and inclination were calculated separately for the binary orbit and for C's orbit with respect to the binary's center of momentum, treating the inner binary as a single companion mass. This does not account for the gas cloud potential; because of this, calculations of initial orbital parameters give seemingly unphysical results such as eccentricities greater than one.

In the second regime, we took systems which had ended resembling the Alpha/Proxima Centauri system today and repeated them, adding in a disk of test particles around Alpha Centauri B. This simulation was also repeated with the third star removed, so that Proxima's effects could be isolated. The parameters used were the same as for the first regime, except that the larger number of particles meant it was infeasible to continue the simulations beyond 10 Myrs. By this point most of the interactions have already taken place and the systems are fairly stable. A total of 224 simulations of this type were performed.

The third regime consists of planet formation simulations in a disk around Alpha Centauri B. Thirty-six simulations were performed over several discrete values of truncation radius, disk slope, and disk density. These simulations represent a later time period, when the three stars have become well-separated and the gas cloud has dispersed, so the third star and the user module forces were omitted, and the wide-binary algorithm was used \citep{2002AJ....123.2884C}. These integrations have shorter time steps (one day) due to the closer object spacing. In {\sc mercury}'s hybrid symplectic mode (as used here), whenever two objects approach each other within a few Hill radii, the algorithm switches to a slower mode which more accurately calculates whether the objects collide. Collisions are simple, perfect accretion. As the detailed physics of collisions are still a rapidly developing field with significant uncertainty, and we are interested in the end state of how the mass groups up rather than intermediary details, we consider this simple scheme appropriate for the scope of this study.

As our computational resources are limited and our interest in planet formation is motivated largely by claims of planets around Alpha Centauri B, our simulations focused on planet formation around B and not A. However, although the two stars are not the same mass, their masses are similar enough relative to the range of masses for planet-hosting stars ($<0.1 $M$\odot$ to $>2.5 $M$\odot$). It is plausible that planet formation processes around Alpha Cen A and B are closely symmetric and any systematic differences due to the stellar mass differences in the underlying planet formation processes are likely to be small (less than a factor of 2) compared to variations in outcomes due to the variation in disk parameters explored. Therefore we expect simulations for formation around one star to be fair predictors of the distribution of likely planets around the other star, within the scope of this study and given the uncertainties in the assumptions.

%------------------------------------------------------------------------------
\subsection{Stellar Mass Cases}

Two different sets of masses were used for the simulations, representing two different scenarios: interactions at early or late times. In the later scenario, the stars have already achieved their current masses, with the binary stars being near a solar mass and Proxima much smaller. The early interaction scenario is following that presented by \cite{2012Natur.492..221R}, in which the stars are still accreting mass when Proxima is ejected, preventing it from accreting as much as its companions. In this case, all three stars have masses of 0.123 M$_{\odot}$. Neither case is intended a perfect representation of the true physical situation, but they provide outer bounds on the strength of the interactions, with the reality likely lying closer to the early case.

%------------------------------------------------------------------------------
\subsection{Orbital Parameters}
Alpha Centauri A and B and Proxima Centauri are represented as Alpha Centauri A, B, and C in the simulations, and may be referred to as such in this paper when talking about the stars in the simulations, as opposed to the actual physical system. 

In {\sc mercury}'s default mode, a single central body is chosen, and the other objects are listed as ``big'' objects with orbital parameters relative to the central body, although internal calculations are carried out in center-of-momentum coordinates. Star A is the largest, and was used as the central object, with Alpha Centauri B in an orbit around A with parameters randomly generated within specified ranges: semimajor axis (a) in the range 23.4-28.1 AU, eccentricity (e) between 0 and 0.52, and zero inclination (i). Parameters were then generated for C so that it had a pericenter between one and ten times the apocenter of B. C could have an initial eccentricity between 0 and 0.75, and any inclination. Both stars were also given random values for the argument of pericenter (g), longitude of the ascending node (n), and mean anomaly (M). These parameter ranges were chosen based on the assumption that the interactions we are interested in would cause the binary to become tighter while Proxima receded, and that both orbits would likely become more eccentric in the process. When close encounters between two objects occur, the simulation switches to a variable-time step regime with slower processing time but greater accuracy, and when objects are far apart it uses a fixed time step, which we set to 10 days. The close-encounter radius used was three Hill radii.

The binary parameters used are based on \cite{2002A&A...386..280P} ($M_A=$ 1.105, $M_B=$ 0.934) and \cite{1999A&A...344..172P} ($a_{AB}=$ 17.57'', $e=0.5179$). Proxima's expected orbital parameters of $a$ near 10,000 AU or above and $e$ just below 1 are based on the arguments in \citep{2006AJ....132.1995W} that it is unlikely for Proxima to be seen so near Alpha Centauri in both position and velocity unless it is bound, and that given that, it is most likely eccentric and near apocenter, when it moves most slowly along its orbit.

Calculation of orbital parameters from the simulations is done instantaneously based on the system's energy and momentum. To calculate the outer third star's orbit, the inner binary is treated as a single mass at the binary's center of momentum. The equations used are as follows:
\begin{equation}
a = \frac{-\mu}{2\epsilon}
\end{equation}
\begin{equation}
e = \sqrt{ 1 + \frac{2\epsilon \bar{h}^2}{\mu^2}} 
\end{equation}
\begin{equation}
i = arccos(\frac{h_z}{\bar{h}}).
\end{equation}
The gravitational parameter $\mu$ is the sum of the masses involved ($M_A$ and $M_B$ for calculations of the binary parameters, and all three stars' masses for the outer star's orbit). The specific orbital energy is
\begin{equation}
\epsilon = \frac{v^2}{2} - \frac{\mu}{r}.
\end{equation}
The angular momentum $h$ is the cross product of the stars' separation $r$ and relative velocity $v$, where $h_z$ is the angular momentum in the $z$ direction, and $\bar{h}$ is the magnitude of the angular momentum.

%------------------------------------------------------------------------------
\subsection{Disk Stability}

By adding disks around each of the stars in the inner binary, we were able to see the influence of the triple-system interactions on protoplanetary disks in Alpha Centauri. Each disk consisted of 100 massless test particles on circular orbits, coplanar with the binary, spaced evenly from 0.1 AU to half of the binary's semimajor axis $a_{bin}$ (typically around 10 AU). We ran pairs of simulations, in which one contained only the inner binary and the other included the third star, to separate out the effect of the third star on the disks.

At each timestep, we tracked whether each particle was still stable (orbit's eccentricity is less than 1 and semimajor axis is within 20\% of initial value) as well as whether it had been removed from the system (i.e. collided with a star or traveled beyond the 100,000 AU ejection radius) and found that stability is a very reliable predictor of removal: objects with orbits outside the above bounds almost always suffered an ejection or collision before the end of the simulation. Due to the large ejection radius used for the simulations, the destabilization time appears to more accurately track when changes to the system happen than the removal time. 

Based on this, we define the truncation radius as the distance from the star which best defines an inner section closest to completely filled and outer section closest to completely depleted. That is, for each particle's orbital radius, we calculate the stable fraction inward ($f_{in}$) and outward ($f_{out}$) of this point, and the truncation radius $r_{tr}$ is defined as the radius at which inward stability is maximized and outward stability minimized, i.e. where $|1-f_{in}|+f_{out}$ is minimized. On timescales shorter than 1,000 years, particles in the outer disk oscillate between stable and unstable, making the truncation radius not especially meaningful until after this time, at which point it tracks the disk shape well.

%- - - - - - - - - - - - - - - - - - - - - - - - - - - - - - - - - - - - - - - 
\begin{figure*}
\begin{center}
\includegraphics[scale=0.9]{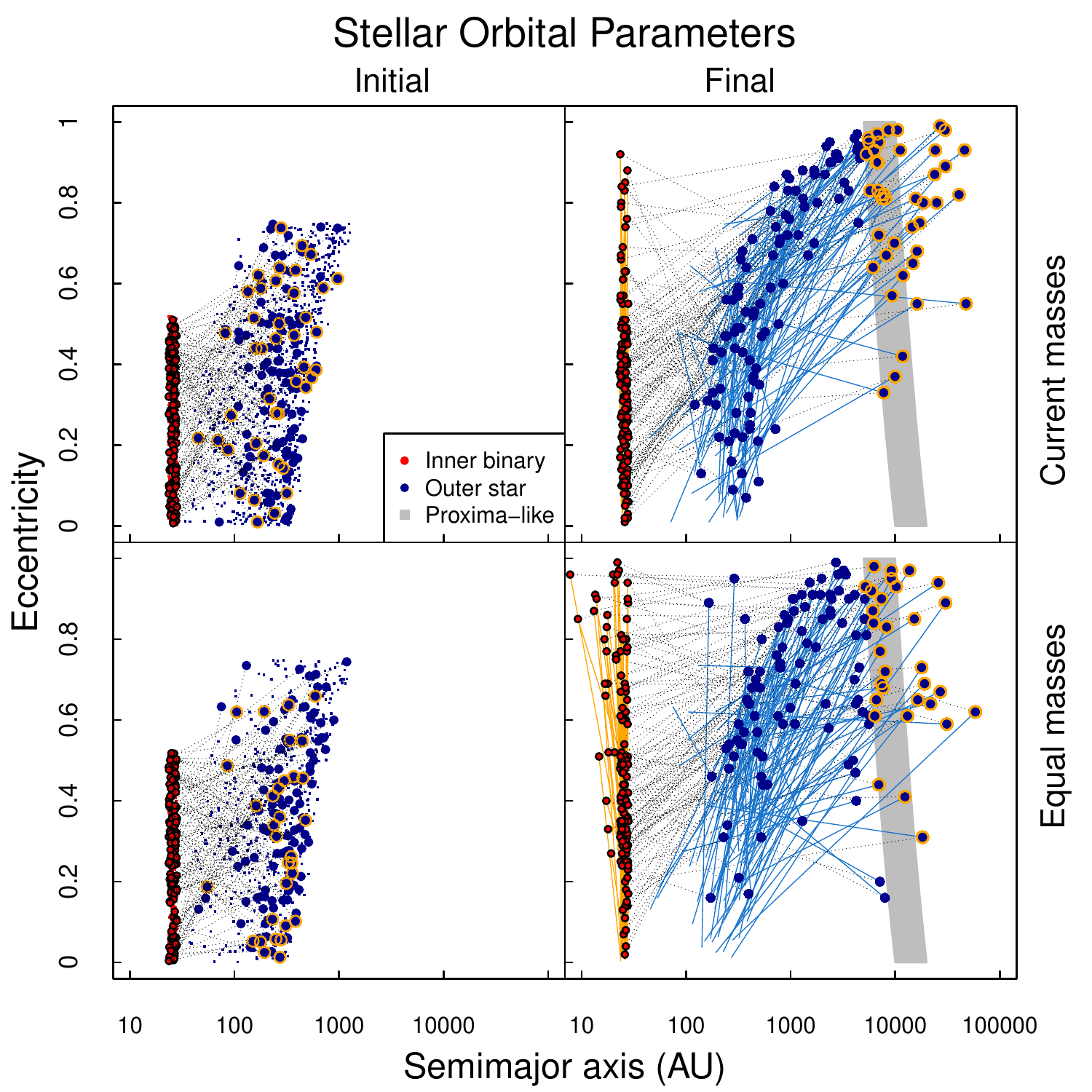}
% created by ?
\caption{Initial (left) and final (right) parameters from 947 simulations using current stellar masses (top) and 575 simulations where all stars have masses of 0.123 M$_{\odot}$ (bottom). Dots in the left panels represent initial parameters for Alpha Centauri B (red) and C (blue). Large dots show simulations which survived, while small ones were disrupted. On the right, final parameters for surviving simulations are marked with large dots, with lines connecting them back to their original parameters. Faint dotted lines connect the stars to their companion from the same simulation. The shaded region shows ``Proxima-like'' orbits with apocenter between 10,000 and 20,000 AU, and orange rings mark system with final orbits with this separation or larger. \label{ae_io}}
\end{center}
\end{figure*}
%- - - - - - - - - - - - - - - - - - - - - - - - - - - - - - - - - - - - - - - 

%%%%%%%%%%%%%%%%%%%%%%%%%%%%%%%%%%%%%%%%%%%%%%%%%%%%%%%%%%%%%%%%%%%%%%%%%%%%%%%
\section{Results and Analysis}

%------------------------------------------------------------------------------
\subsection{Stellar System Simulations} \label{StarSimResults}

The results from suites of simulations of the stars' dynamical interactions can be seen in Fig.~\ref{ae_io}. The left panels show the initial semimajor axis and eccentricity for both B and C. On the right, the final parameters of surviving systems are shown, along with a colored line connecting the points back to their initial parameters, and thin black dotted lines connecting B and C from the same simulation. The shaded region indicates the orbits with an apocenter of between 10,000 and 20,000 AU, which is our criterion for a system to be dubbed ``Proxima-like.'' Systems with an orbit this size or larger are marked on the plot with orange rings. This population is examined with further simulations in the next section. 

The majority of systems are disrupted (small dots in left hand panel). In most of the surviving systems, star B's semimajor axis decreases a small amount and the eccentricities of both B and C increase. The initial orbital conditions do not appear to predict whether a system will survive or become Proxima-like.

%- - - - - - - - - - - - - - - - - - - - - - - - - - - - - - - - - - - - - - - 
\begin{table}[tbp]
\begin{center}
\caption{Outcomes of Simulations \label{simstats} }
\begin{tabular}{l|rr}
\hline
 & Late & Early \\ 
  \hline
Total  &     &     \\
Number & 947 & 575 \\
  \hline
  Survived (\%) & 14.4 & 24.3 \\ 
  Expanded (\%) & 13.3 & 22.6 \\ 
  Proxima-like (\%) & 4.44 & 5.94 \\ 
  Huge orbit (\%) & 0.634 & 2.64 \\ 
  A Ejected (\%) & 0.739 & 13.8 \\ 
  B Ejected (\%) & 0.211 & 8.2 \\ 
  C Ejected (\%) & 84.5 & 52.6 \\ 
  Collision (\%) & 0.211 & 0.696 \\ 
  Broken* (\%) & 1.9 & 16 \\ 
  \hline
\end{tabular}
\tablecomments{Percentages from each simulation type which saw various types of outcomes: survived (no ejection in $10^9$ years); expanded (semimajor axis increased); Proxima-like (C's final orbit had an apocenter $> 10,000$ AU); Huge orbit (C's final orbit larger than 100,000 AU and was counted as ejected, but appeared stable at that point); had a star ejected; saw a collision between stars; or contained output errors due to particularly fast ejection. Simulations started with current masses ($m_A=1.105$, $m_B=0.934$, and $m_C=0.123$ M$_{\odot}$) or all three equal to 0.123 M$_{\odot}$. \newline \hspace*{1.5em} *See Section~\ref{StarSimResults}}
\end{center} 
\end{table}
%- - - - - - - - - - - - - - - - - - - - - - - - - - - - - - - - - - - - - - - 
\begin{table}[tbp]
\begin{center}
\caption{Median change in orbital parameters \label{medians} }
\begin{tabular}{l|rr}
\hline
        & Late    & Early   \\ 
  \hline
  $a_B$ & -0.045  & -0.035  \\ 
  $e_B$ & 0.027   & 0.354   \\ 
  $i_B$ & 13.0    & 37.25   \\ 
  $a_C$ & 9695.87 & 8852.03 \\
  $e_C$ & 0.4     & 0.4805  \\ 
  $i_C$ & -9.8    & -4.7    \\ 
\hline
\end{tabular}
\tablecomments{Median change observed in orbital parameters of simulations which resulted in Proxima-like systems. Assuming the binary orbital parameters changed by the amount seen in typical early-stage interactions, the current parameters of $a_B=23.7$ AU and $e_B=0.5179$ imply initial parameters with a similar semimajor axis and an eccentricity of 0.16. This implies the pericenter would have changed from approximately 19.8 AU to the current 11.4 AU.}
\end{center} 
\end{table}
%- - - - - - - - - - - - - - - - - - - - - - - - - - - - - - - - - - - - - - - 

Table~\ref{simstats} shows the types of outcomes as percentages of the total number of simulations of each type. Comparing the two stellar mass cases, the rate of surviving systems and Proxima-like systems are similar for both the current and equal mass cases, but the fates of disrupted systems differ. In the case where all three stars had small masses, it was far more likely for one of the central stars (A or B) to be ejected, whereas in the other mass case, the outer star's smaller mass made its ejection much more likely. In addition, an error in {\sc mercury}'s output caused by a very high kinetic/potential energy ratio  occurred far more often in equal-mass cases, indicating higher-energy ejections. {\sc mercury} does not save orbital parameters directly, but stores the value $f_v = \frac{1}{1+2(K/U)^2}$ (where $K$ is the kinetic energy and $U$ is the binding energy) in a condensed format of eight 224-bit characters, from which {\sc element} can later calculate them. This allows for a precision of $224^{8-1} = 2.83 \times 10^{-10}$ in $f_v$, so for $K/U$ ratios of $\gtrsim 42,000$ it experiences underflow error and may output $f_v = 0.0$, despite this never being correct for a physical system. This causes the calculations for orbital parameters to return infinity, NaN, or 0. Although the internal values remain unaffected and the integration continues successfully, some time-series information can be lost in cases where the ratio of the star's kinetic to potential energies is particularly large. This only affects stars which are being rapidly ejected, and thus is not significant for the results of our study, which is primarily concerned with the bound systems.

It seems that for most systems with the current stellar masses, the inner stars' orbit was only slightly altered; $a$ decreased by only a fraction of an AU in most cases, while eccentricities and inclinations sometimes remained similar and sometimes increased significantly. For the case in which all three stars have small, equal masses, the changes in orbital parameters are more pronounced, particularly in eccentricity increases. The median changes in orbital parameters are shown in Table~\ref{medians}.

Typically, the triple system remains clearly hierarchical as intended, with the outer star well-separated from the inner system. In a few cases this assumption did not hold, which brings the system into a regime where the integrator used is not accurate. These systems were ignored. This implies that the region of parameter space in which the stars do not remain well separated is unexplored, but this appears to be an unlikely way to form the system in question and does not significantly skew our results.

\begin{figure}[htbp]
\includegraphics[scale=0.65]{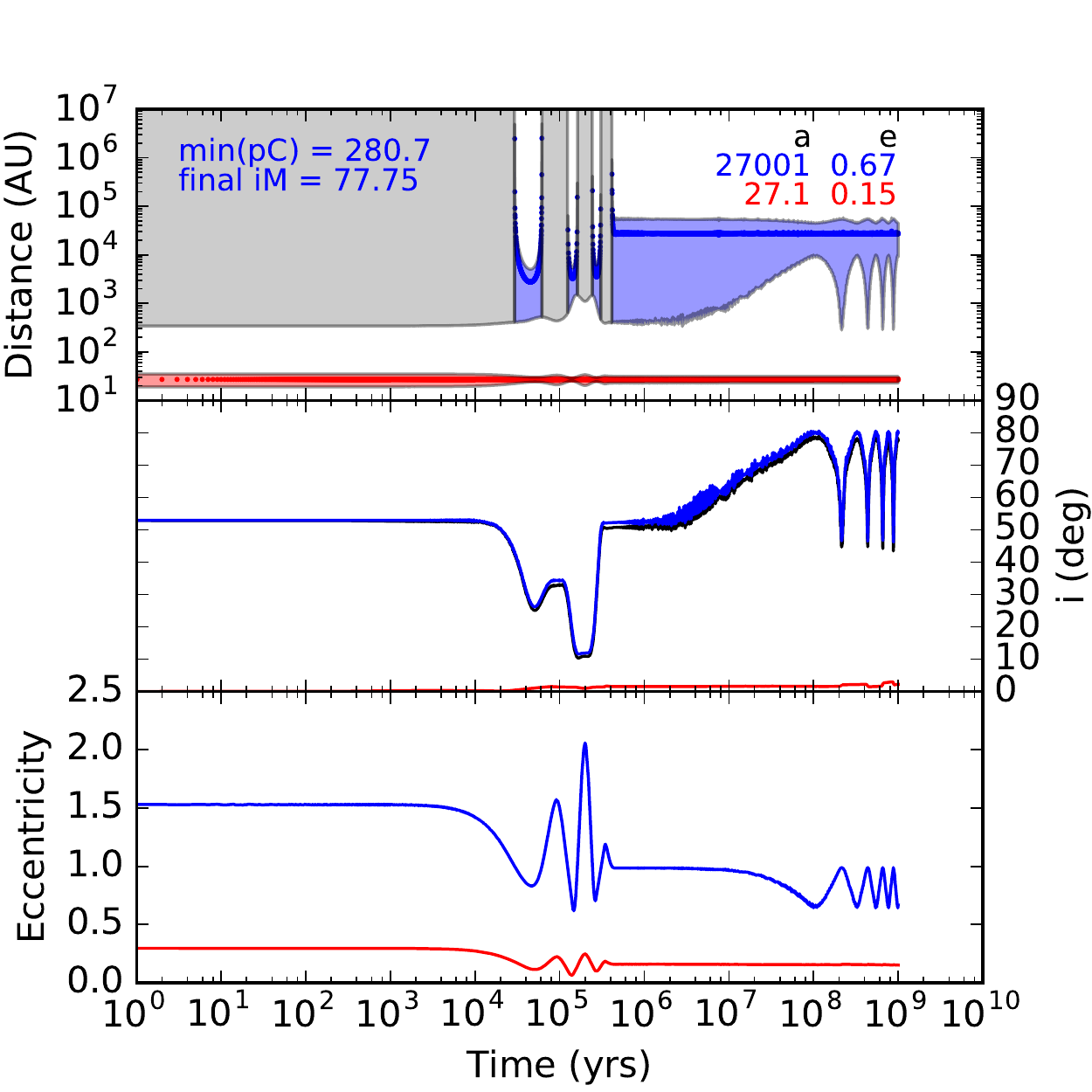}
% figure created by Sims/PlotTimeData.py
\caption{The evolution of a representative triple star system. Inner binary system parameters are shown in red, while the outer star's orbital parameters relative to the inner binary's center of mass is shown in blue. \textbf{Top:} Points indicate semimajor axis over time, with shading extending from pericenter to apocenter. For the outer star, due to the gas potential the calculated orbit sometimes appears unbound, in which case the area from the pericenter up is shown in gray. \textbf{Middle:} System inclinations relative to the binary's initial inclination, and their difference (black). \textbf{Bottom:} Eccentricities. \label{paths}}
\end{figure}

The evolution of the system to a Proxima-like structure generally followed one of two paths. Most commonly, the system began in a state that quickly moved in and out of stability as the gas cloud dispersed, then froze in at a large stable orbit as the gas potential finally disappeared after 440,000 years, consistent with the scenario in which interactions occur early on; an example system is shown in Fig.~\ref{paths}. (Because of the gas cloud potential, initial calculations of the outer binary's eccentricity are incomplete, which is why the eccentricity sometimes appears to be greater than zero initially.) Less commonly, the system remained tightly bound but subject to significant Kozai-Lidov oscillations or general three-body interactions which pushed it into a large orbit at late times (10-100 Myr). We also observe slow ($\approx$ 1 Gyr) Kozai-Lidov oscillations between the eccentricities and inclinations of the inner binary and outer star, causing a gradual oscillation in their pericenters, which may be significant for disk and planet stability.
%- - - - - - - - - - - - - - - - - - - - - - - - - - - - - - - - - - - - - - - 
%- - - - - - - - - - - - - - - - - - - - - - - - - - - - - - - - - - - - - - - 
%%%%%%%%%%%%%%%%%%%%%%%%%%%%%%%%%%%%%%%%%%%%%%%%%%%%%%%%%%%%%%%%%%%%%%%%%%%%%%%
\subsection{Disk Simulations} \label{disks}

We performed simulations of coplanar test particle disks around Alpha Centauri A and B, in pairs of simulations using the same initial conditions in all ways except that one included Proxima Centauri while the other did not. These simulations indicate the region of stability around each star in which protoplanetary disk material can remain throughout the star's lifetime, allowing for later planet formation. The initial conditions for these systems were copied from the simulations that produced Proxima-like systems in the previous section. Due to the extremely sensitive, chaotic nature of dynamics and the finite precision of floating point number computation, sometimes the simulations resulted in different final orbits from the original version. Therefore, the following analysis will make a distinction between when all simulations are being discussed, or only those which resulted in Proxima-like final configurations.

\begin{figure}[htbp]
\includegraphics[scale=0.6]{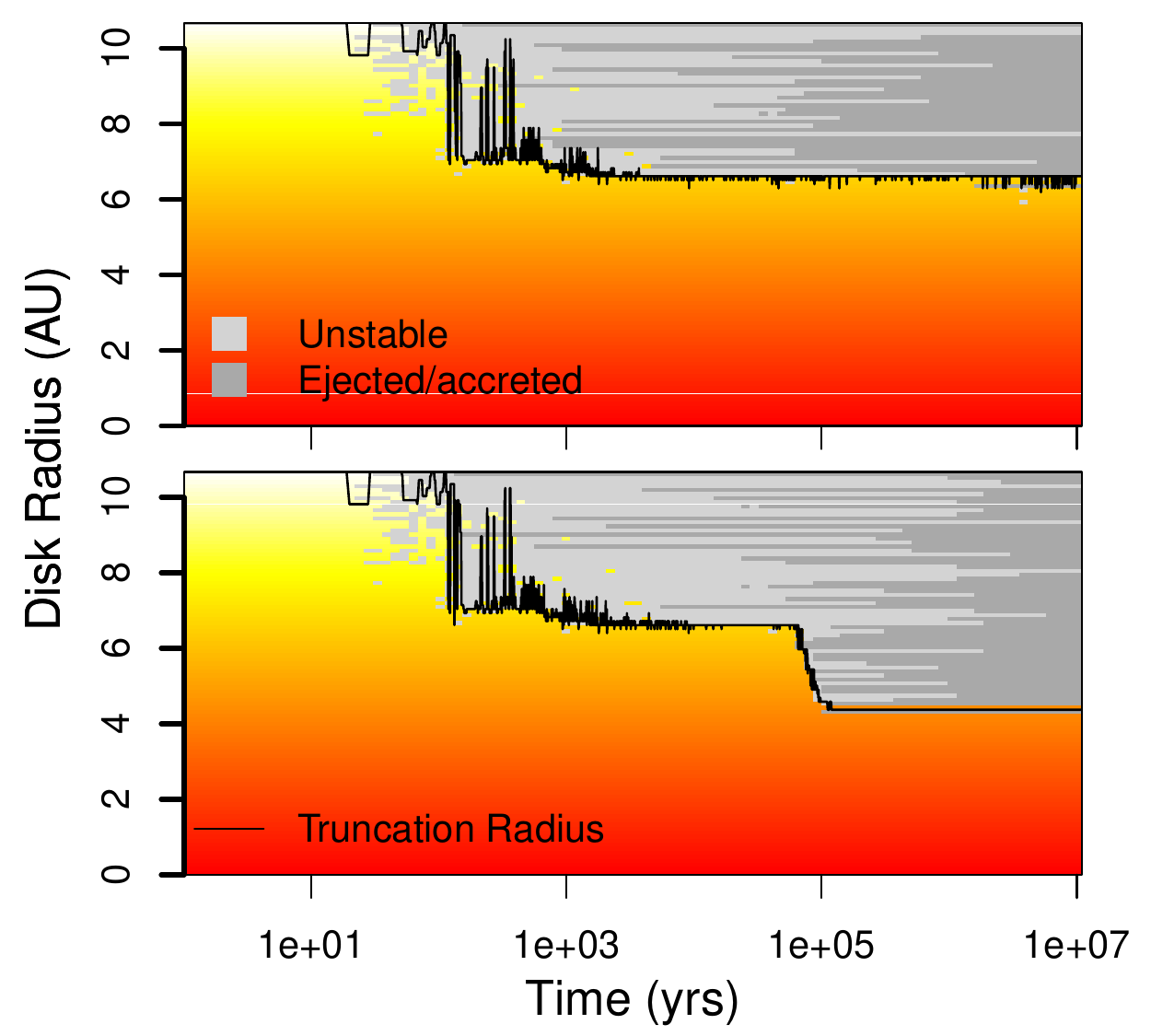}
% created by Analysis/ReadDisk.R -> ReadDiskPlot4.R
\caption{Disk survival for a typical early-interaction Proxima-like simulation of a disk around Alpha Centauri B, for (top) the binary, and (bottom) the triple. The presence of the third star typically removes about 1-2 AU of the disk in the equal mass case. Disk material within the truncation radius remains undisturbed. \label{diskimg}}
\end{figure}

Generally, we find that the disk particles are truncated at some distance from the star, which we call the truncation radius $r_{tr}$. The edge is clean, in that almost all particles' orbits within this radius remain stable, while almost all particles beyond this point are removed from the simulation through ejection or collision with a star. The behavior of disks around the two stars in the binary is very similar, with a slightly larger disk remaining around the slightly larger primary star (companion mass ratio $= 0.85$). Typically, in simulations without the third star, the outer portion of the disk was destabilized within about 10,000 years, while disk material within a certain truncation radius ($r_{tr}$) remained stable for the 10 Myr duration of the simulations. The presence of the third star induced additional truncation of one or two AU at around 100,000 years, as the binary's pericenter shrinks (see Fig.~\ref{diskimg}). 

%- - - - - - - - - - - - - - - - - - - - - - - - - - - - - - - - - - - - - - - 
\begin{figure}[htbp]
\includegraphics[scale=0.8]{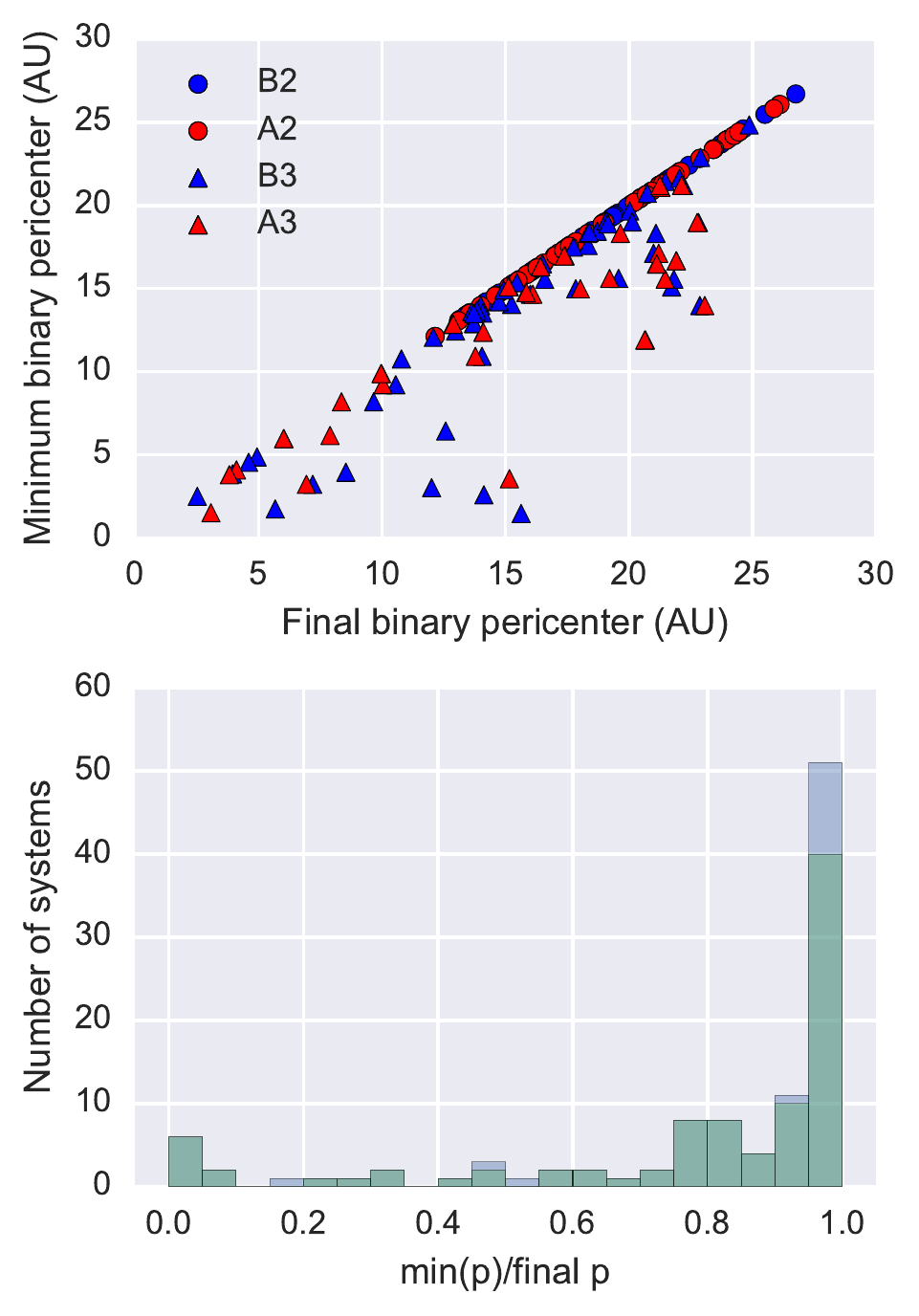}
% old version created by Sims/AnalyzeTimeData.R
% current version created by Sims/ReadDiskSum.py
\caption{(a) Minimum binary pericenter vs. final pericenter for each disk simulation. The legend indicates the simulation group, where ``B2'' indicates two-star simulations with a disk centered on star B, ``B3'' indicates a similar three-star system, etc.
(b) Histogram of the inner binary's minimum pericenter in simulations which did (green) and did not (blue) produce Proxima-like systems, as a fraction of the final pericenter. \label{PeriHist}}
\end{figure}
%- - - - - - - - - - - - - - - - - - - - - - - - - - - - - - - - - - - - - - - 
%- - - - - - - - - - - - - - - - - - - - - - - - - - - - - - - - - - - - - - - 
\begin{figure}[htbp]
% created by Sims/AnalyzeTimeData.R
\includegraphics[scale=0.6]{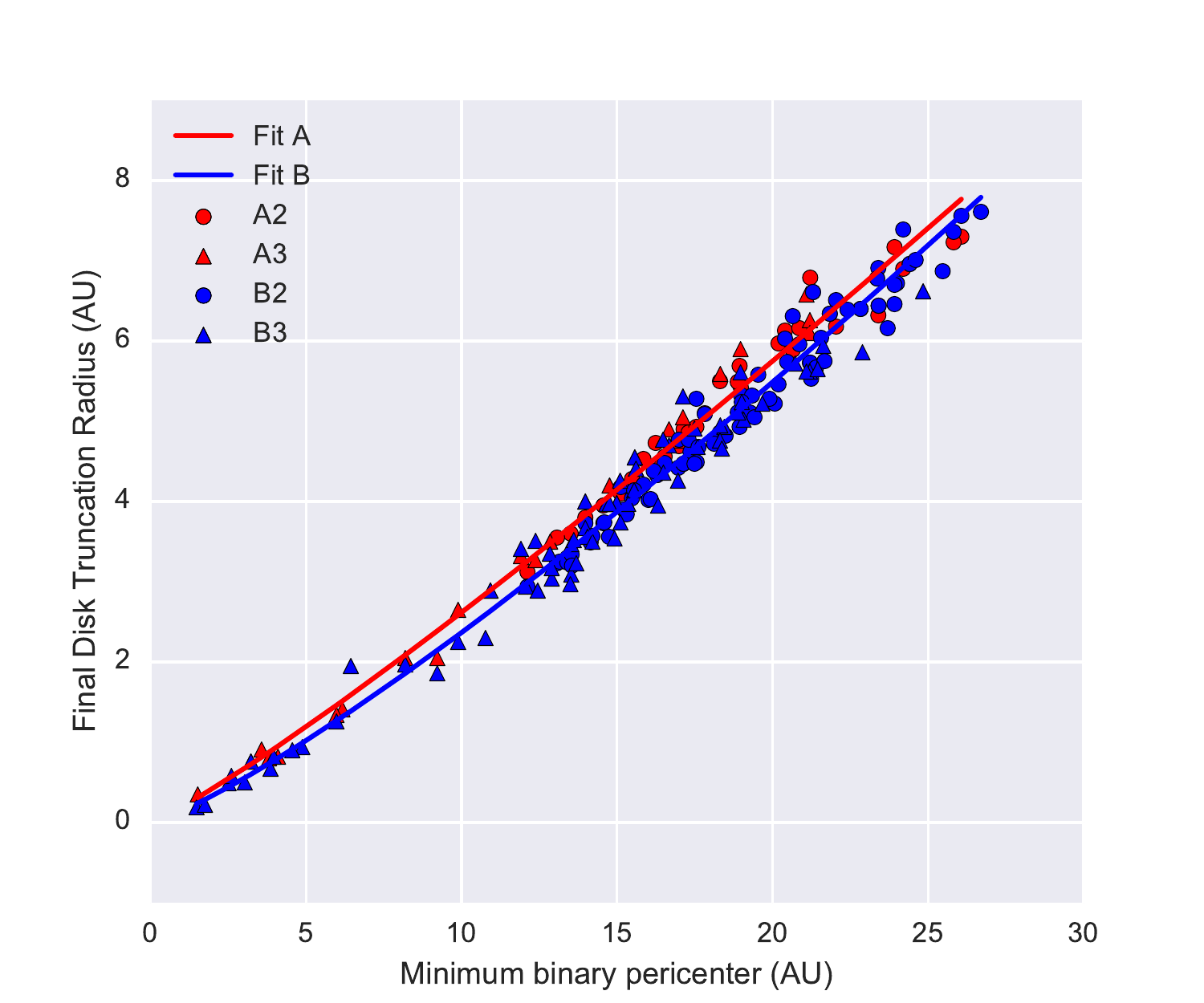}
% created by Sims/ReadDiskSum.py
\caption{Truncation radius vs. minimum pericenter. As in Fig.~\ref{PeriHist}, ``A2'' indicates a two-star simulation with a disk centered on star A, etc. Red points show disk simulations centered on Alpha Centauri A, while blue are B. Circles show simulations including only the inner binary stars, while triangles contain Proxima as well. The lines show power-law fits to the points of the same color. Outlier points identified in Fig.~\ref{outliers} were excluded from this plot and the fits. Large points indicate simulations which resulted in a Proxima-like orbit. The coefficients of the fits are shown in Table~\ref{fitcoefs}. \label{rtrfit}}
\end{figure}
%- - - - - - - - - - - - - - - - - - - - - - - - - - - - - - - - - - - - - - - 
\begin{table}[tbp]
\begin{center}
\caption{Truncation radius model coefficients \label{fitcoefs} }
\begin{tabular}{l|rrr}
\hline
Subset & $a$      & $b$      & $R^2$    \\ 
  \hline
A      & 1.14 & 0.192 & 0.9926 \\
B      & 1.21 & 0.145 & 0.9862 \\ 
Both   & 1.17 & 0.168 & 0.9839 \\ 
\hline
\end{tabular}
\tablecomments{Coefficients for a power law model predicting disk truncation radius $r_{tr}$ as a function of minimum binary pericenter $p_{min}$, of the form $r_{tr} = b \times (p_{min})^a$. The leftmost column indicates which subset of the data the model was fit to: disks centered around Alpha Centauri A or B, or both sets combined. The fits are shown in Fig.~\ref{rtrfit}. The score column shows the model's coefficient of determination.}
\end{center} 
\end{table}
%- - - - - - - - - - - - - - - - - - - - - - - - - - - - - - - - - - - - - - - 
\begin{figure}[htbp]
%\includegraphics[scale=0.6]{Rtr_vs_minP.pdf}
% created by Sims/AnalyzeTimeData.R
\includegraphics[scale=0.8]{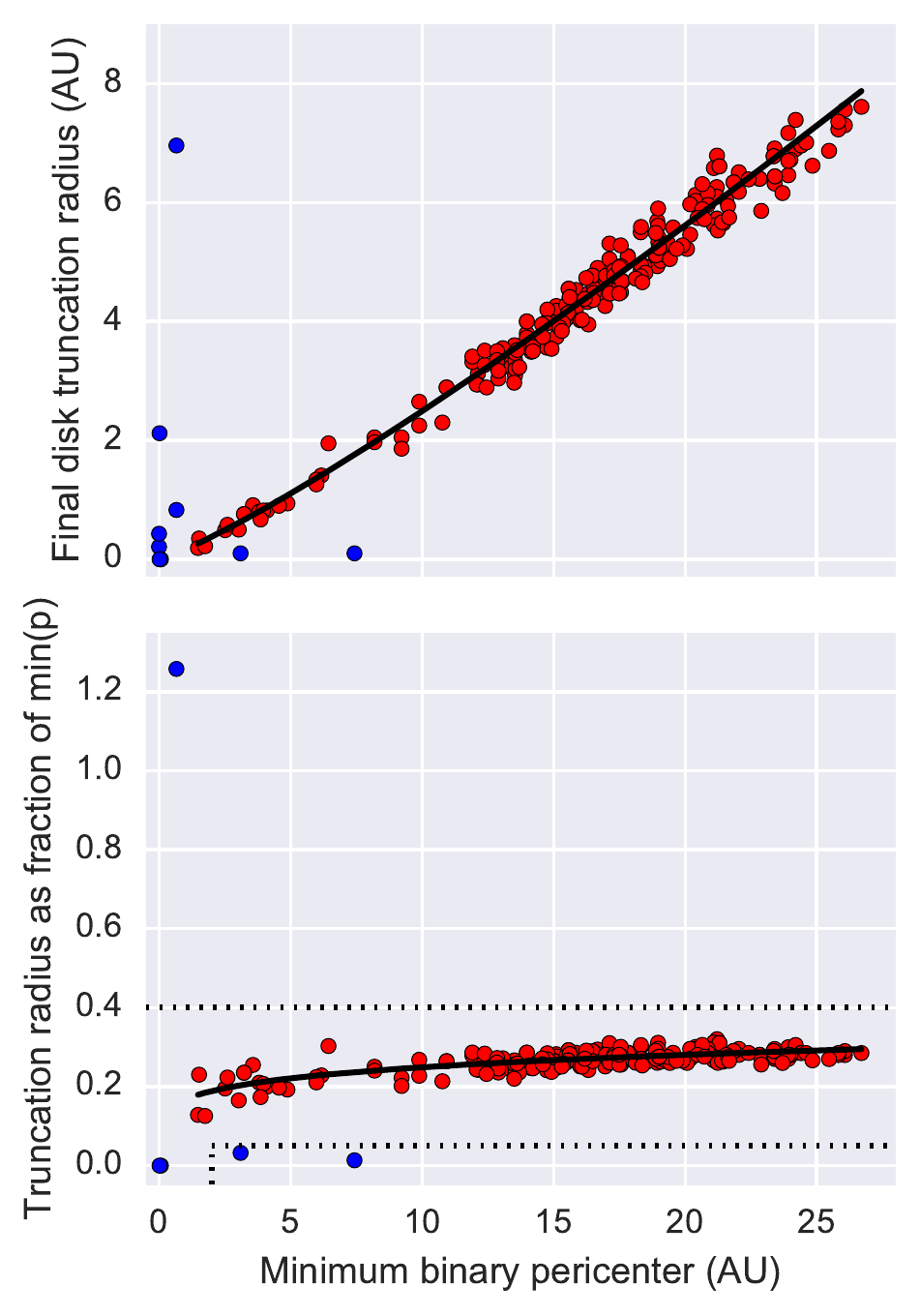}
% created by Sims/ReadDiskSum.py
\caption{(a) Truncation radius vs. minimum pericenter. The red points are considered part of the main trend, while the blue points are outliers. The black line shows the fit to all disk simulations (``Both'' in Table~\ref{fitcoefs}). (b) The above data with truncation radius as a fraction of the minimum pericenter. The dotted lines show the boundaries used to exclude outliers. Two more outlying points, with pericenters close to zero and truncation radii many times larger, lie far off the vertical scale, and another two are unplottable because their $p_{min}$ values round to zero. \label{outliers}}
\end{figure}
%- - - - - - - - - - - - - - - - - - - - - - - - - - - - - - - - - - - - - - - 

Although previous literature commonly discusses disk extent as a fraction of the binary's semimajor axis, we find the minimum pericenter $p_{min}$ (the smallest observed pericenter over the course of the simulation, or min($a(1-e)$) where $a$ and $e$ are the binary's instantaneous semimajor axis and eccentricity), to be the most relevant parameter and will primarily discuss the disk extent in relation to it. 
The binary's closest approach, while probably an even better predictor, cannot be reliably measured because the orbital parameters are output at a finite number of timesteps which often miss the closest approach. In most cases, minimum pericenter closely tracks closest approach, but because it is based on instantaneous parameters, it sometimes gives odd results in quickly-evolving systems (see the discussion of outliers below).

Ultimately, the binary's minimum pericenter over the course of the simulation is the best predictor we have of the final disk truncation radius, with a Pearson correlation coefficient $R = 0.95$. Exclusion of the seven outlier points, discussed more below, improves this correlation to 0.987. Of the directly observable system properties, final binary pericenter $p_f$ (instantaneous pericenter at the end of the simulation) is also a good predictor ($R = 0.91$), as the final and minimum pericenters are themselves strongly correlated ($R = 0.90$). 

The distributions in these simulations are non-Gaussian, so rather than using the mean and standard deviation to characterize ranges of values, we use the more robust median and median absolute deviation (MAD). The median radius at which the disk was truncated in simulations with the Alpha Centauri binary alone, in units of the minimum binary pericenter $p_{min}$, was $0.281p_{min} \pm 0.014$ for disks around Alpha Centauri A, and $0.267p_{min} \pm 0.016$ for B. The minimum pericenter in simulations with a third star was a median of $18.3\% \pm 26.5\%$ smaller than binary simulations with the same initial orbits, resulting in a corresponding reduction in disk radius. The outcomes for $r_{tr}$ in triple systems covered the full possible range, from disks the same size as the binary-only simulation, to 100\% loss of disk material. The relationship between minimum and final pericenter is shown in Fig.~\ref{PeriHist}a and the distribution in Fig.~\ref{PeriHist}b. The fractional truncation radius in triple systems is similar to the binary case, but slightly smaller and with a larger spread: $0.274p_{min} \pm 0.030$ and $0.264p_{min} \pm 0.027$ for disks around Alpha Centauri A and B respectively.

The main trend between truncation radius and minimum pericenter can be fit very well with a power law using an exponent slightly above 1, as shown in Fig.~\ref{rtrfit}. Coefficients for a power law model fit to disks centered on Alpha Centauri A or B, and on the combined dataset, are shown in Table~\ref{fitcoefs}. There were 150 simulations around Alpha Centauri B and only 74 around A due to logistical limits on computation time and greater observational interest in B, so in the combined fit the A samples were weighted proportionally higher. A similar trend is seen around each star, with slightly larger disks remaining around the more massive one, as is expected. 

This is consistent with previous studies by \cite{1999AJ....117..621H}, which examined stability in coplanar binary disks for varying eccentricities. According to their Table 4, they found ``critical semimajor axes'' (equivalent to our ``truncation radius'') of 2.79 and 2.54 AU around Alpha Centauri A and B, respectively. Our estimates of 3.0 and 2.7 AU for the current pericenter of 11.2 AU are roughly similar, though somewhat higher; the differences are likely due to their assumption that the truncation radius as a fraction of semimajor axis does not vary with semimajor axis, where in our model it does, and possibly also due to the different integrators used.

%critical a = 2.79 AU for A, 2.54 AU for B

%A: 0.192 x (11.2)^1.14 = 3.0
%B: 0.145 x (11.2)^1.21 = 2.7

%.29  * 23.4 = 
%.258 * 23.4 = 

Although the power law model fits the data very well in the studied parameter range, because the exponent is larger than one, the predicted disk radius as a fraction of the pericenter can become arbitrarily large as the binary orbit grows in size, and could eventually surpass the binary orbit itself, which is clearly unphysical. Therefore, we expect the relationship must become linear or otherwise change at large separations; further study is needed before extrapolating to larger binary orbits. Additionally, the model coefficients' apparent dependence on mass should be studied at more stellar masses and mass ratios.

All but nine of the 224 stable disk simulations have a $r_{tr}/p_{min}$ value lying between 0.05 and 0.4. These outlying points are identified in blue in Fig.~\ref{outliers}a and b. Seven points appear to be outliers from the main trend, and are omitted based on the cuts shown in the dotted lines of Fig.~\ref{outliers}b, which were chosen by eye. The five systems lying above this limit do so because their binary pericenters drop to nearly zero for a very brief time (less than an orbit) while the gas cloud is dispersing, and the stars do not actually pass this close to each other before continuing to be perturbed into another orbit, so the disk is not significantly perturbed. Two points also lie below the trend, with truncation radii of 0.1 AU but minimum pericenters of several AU; these disks contain gaps at small radii that confuse the disk fitting algorithm. The other two blue points lie close to the origin, and are likely a continuation of the downward curve seen for small-pericenter systems. However, because their $r_{tr}$ values round to zero, they are excluded along with the outliers because they must be omitted from the fitting algorithm, which requires taking $log_{10}$ of all $r_{tr}$ and $p_{min}$ values. These nine simulations represent 4\% of the three-star systems. When discussing the overall trend of disk survival these special cases are omitted. However, they are possible, though rare, outcomes of multistellar systems, and readers should remain aware that exceptions to the general trend exist.

The minimum binary pericenter is the most important factor in determining the amount of disk material available for planet formation. When we observe a stellar system, we can only observe the pericenter it has now, without knowing how much smaller it might have been in the past and thus how much additional truncation may have occurred. However, most systems had minimum pericenters which were similar to or only slightly smaller than their final pericenters -- the median minimum pericenter was 0.932 times the final pericenter. For the Alpha Centauri binary, with a current pericenter of 11.4 AU, this means we could expect the minimum pericenter the system has experienced to be around 10.6 AU. In our model for truncation radius as a function of pericenter, these values would imply disk truncation radii of 2.8 and 2.5 AU, respectively. Obviously, the possible values for minimum pericenter and truncation radius extend down to zero, and properties from the past are no longer observable. Our goal is to find the most probable values.

The final truncation radius for the disk is closely correlated with the pericenter of the inner binary, and since the change in semimajor axis is generally small, the degree of truncation is primarily a tracer for the amount by which the third star increases the eccentricity of the inner binary. Systems in which the binary's size was considerably reduced necessarily had smaller disks, but those with larger semi major axes could have disks that are either large or small, depending on their eccentricity.

For systems which end up similar to Proxima Centauri, the precise parameters of the outer star were not predictive of any other simulation parameters, including inner binary orbit or disk stability. The two systems with particularly large $a_C$ did see less truncation, but there was no statistically significant correlation. It appears Proxima does not directly interact with the stability of disks in the system, but rather induces changes in the binary, which acts as an intermediary. 

The above numbers assume the more likely situation, in which triple-system interactions take place early in the stars' lifetimes (that is, when all three stars are still small). This means that Proxima's influence on the binary stars is larger compared to late-stage interactions in which they have accreted more mass. In this less physically plausible case, the inner binary's orbit changes by a relatively small amount; the median pericenter reduction from the two- to three-star case was $3.7\% \pm 4.9\%$.

\begin{table*}[tb]
\begin{center}
\caption{Total Dust Mass in Truncated Disks (M${\oplus}$) \label{dustmass}}
\setlength{\extrarowheight}{1pt}
\begin{tabular}{ll|rrr|rrrr}
 & \multicolumn{1}{r|}{\boldmath$r_{tr}$} &      & 2.5  &       &       &  2.8 &      & AU \\
\hline
\boldmath$\alpha$ & \diagbox[]{\boldmath$\sigma_0$}{\boldmath$r_{ice}$} & 2.5  &  2.7  &  3.0  &  2.5  &  2.7  &  3.0 & AU \\
%     &      &      &       &       &       &       &       \\
\hline
    & 0.3 &  0.39 & 0.37 & 0.37 & 0.56 & 0.45 & 0.41 & \\
 1  & 1   &  1.17 & 1.11 & 1.11 & 1.67 & 1.34 & 1.23 & \\
    & 3   &  3.52 & 3.33 & 3.33 & 5.00 & 4.02 & 3.68 & \\ \hline
    & 0.3 &  1.16 & 1.12 & 1.12 & 1.49 & 1.27 & 1.20 & \\
3/2 & 1   &  3.49 & 3.35 & 3.35 & 4.48 & 3.81 & 3.59 & \\
    & 3   & 10.46 & 10.05 & 10.05 & 13.45 & 11.44 & 10.76 & \\
\end{tabular}
% calculated from Sims/JangCondell.py
\tablecomments{Mass of solids in disk, given various combinations of the disk properties $\alpha$, $\sigma_0$, $r_{ice}$, and $r_{tr}$. The disk mass can range from less than one to more than 10 M${\oplus}$ (compare to the mass of Neptune, 17.1 M${\oplus}$), implying that the presence or absence of terrestrial planets depends heavily on the parameters of protoplanetary disks. Note that for $r_{tr} = 2.5$ AU, changing the ice line from 2.7 to 3.0 AU has no effect, and the values for $r_{ice} = 2.5$ AU differ slightly only because of rounding errors.}
\end{center}
\end{table*}

%------------------------------------------------------------------------------
\subsection{Implications for Planet formation} \label{pltform}

The effects of any potential past interactions between Alpha and Proxima Centauri will have resulted in the orbital parameters we see today. The most likely interaction based on our simulations was an increase in eccentricity in the binary, resulting in a smaller pericenter and removal of an outer portion of the protoplanetary disk. This means that at one time there likely was more material in the disk, but it would have been removed well before the epoch of planet formation. The only clear signature we predict would be a slightly smaller disk than if the binary had formed in isolation. Assuming disks of these sizes, we consider expected planet formation in each case, first through an analytic method, then with N-body simulations of planet formation.

%------------------------------------------------------------------------------
\subsubsection{Planet formation: Jang-Condell Model} \label{pltformmodel}

\cite{2015ApJ...799..147J} proposes an analytic method to predict the ease of planet formation in binary star systems based on the amount of disk material remaining, with additional background found in \cite{2007ApJ...654..641J, 2008ApJ...683L.191J}. The system is based on the number of simulations (out of a set of 18 total) retaining sufficient mass to form giant planets via the disk instability or core accretion methods ($N_{DI}$ and $N_{CA}$). An equation is then fit to the cases explicitly simulated, which allows one to predict how many simulations would allow planet formation based on the masses ($\mu$ and $M_*$) and orbital parameters ($a$ and $e$) of the system. This equation is reproduced here:
\begin{equation}
N = \sum_{i,j,k,l} c_{ijkl} \Big(\frac{a}{1 AU}\Big)^i e^j \mu^k \Big(\frac{M_*}{M_\odot}\Big)^l
\end{equation}
where the values of $c_{ijkl}$ are found in Table~3 of \cite{2015ApJ...799..147J}.

For Alpha Centauri, out of 18 simulations with a variety of disk parameters, three retained enough dust to form a solid giant planet core (i.e. $N_{CA}=3$), compared with seven or higher for systems with confirmed planets. This puts the Alpha Centauri binary at the lower edge of the parameter space which allows for planet formation. Inferring the most likely initial parameters for Alpha Centauri by subtracting the median change in $a$ and $e$ in our simulations from its current parameters (giving $a_0=23.5$ AU, $e_0=0.16$) gives an analytic $N_{CA}$ estimate of 11.2 based on Eqn.~1, implying giant planets would form easily in the pre-truncation disk; however, truncation occurs at approximately 100,000 years, significantly before core accretion is believed to occur. The pre-truncation disk instability parameter $N_{DI}$ remains unfeasibly low at 1.89.

For the sake of estimating the likelihood of terrestrial planet formation, we follow the method of \cite{2015ApJ...799..147J} in calculating the total dust mass likely to remain in the extent of the stable disks. Using the Minimum Mass Solar Nebula (MMSN) \citep{1981PThPS..70...35H} as a point of comparison, we assume a power law density profile
\begin{equation}\label{SigProfile}
\Sigma(r) = \Sigma_0 \Big(\frac{r}{r_0}\Big)^{-\alpha}
\end{equation}
where $\Sigma(r)$ is the surface density at $r$, the distance from the central star; $\alpha$ is some exponent determining the slope; $\Sigma_0$ is a density normalization parameter; and $r_0=1$ AU. We can find the total disk mass $M_{tot}$ for a given set of disk parameters (surface density normalization $\sigma_0$, exponent $\alpha$, and ice line $r_{ice}$) by integrating from the inner disk edge $r_i$ to the truncation radius $r_{tr}$.

\begin{equation}
M_{tot} = \int_{r_i}^{r_{tr}} \Sigma(r) 2 \pi r dr \\
\end{equation}
For $\alpha=3/2$ and constant $\Sigma_0$, this simplifies to
\begin{equation}
M_{tot}(r_{tr}) = 4 \pi \Sigma_0 {r_0}^{3/2} (r_{tr}^{1/2} - r_i^{1/2}) \\
\end{equation}
while for $\alpha=1$ it comes to
\begin{equation}
M_{tot}(r_{tr}) = 2 \pi \Sigma_0 r_0 (r_{tr} - r_i). \\
\end{equation}

The disk density changes significantly at the system's ice line, $r_{ice}$, and each section must be integrated separately. Water inside this distance is photo-evaporated, while further out it remains frozen and contributes to the solid mass of the disk. The Hayashi values for $\Sigma_0$ inside and outside of the ice line are 7.1 and 30~g/cm$^2$ respectively. 

The physical parameters of protoplanetary disks are still uncertain, so we consider a range of plausible values in four different parameters: $r_{tr}$, $\alpha$, $\sigma_0$, and $r_{ice}$. The values of $r_{tr}$ represent disks formed in systems with and without three-star interactions, as described in the previous section. The potential range of disk densities is not yet well understood, so we multiply the density normalization $\Sigma_0$ by a scale factor $\sigma_0$ between 0.3 and 3, scaling the total density relative to the MMSN. Canonically, $\alpha=3/2$ and $r_i = 0.35$ AU. Although the 3/2 slope value is favored in theoretical disk calculations, observations tend to favor $\alpha=1$; we use both values here. For $\alpha=1$ cases, we scale the density $\Sigma_0$ such that the total mass in disks out to 36 AU would be equal. The location of the ice line is not precisely determined, and so we use a range of values (2.5, 2.7, and 3.0 AU) that encompass commonly used Solar values. Alpha Centauri B is slightly smaller and cooler than the Sun, while A is slightly larger and hotter, so we assume that the variation in ice line location with stellar mass is small compared to the uncertainty in location we have already included. Additionally, we assume that the stars' separation is sufficiently large that they do not strongly affect each others' ice line locations.

Using these values, we calculate the total dust mass of the disk that will remain stable after interactions remove the outer mass. The values are shown in Table~\ref{dustmass}. The total dust mass values range from less than 0.1 to more than 10 M$_\oplus$. Assuming canonical physical properties and efficient accretion, we expect Alpha Centauri B to have formed several terrestrial planets within a few AU, indicating a significant probability for planets in the habitable zone. The density normalization $\sigma_0$ is the most significant parameter for determining the likelihood of planet formation,, followed by the slope $\alpha$. The other parameters ($r_{tr}$ and $r_{ice}$) have relatively minor effects.

%------------------------------------------------------------------------------
\subsubsection{Planet formation: Planetesimal Disk Simulations} \label{pltformsims}

We performed simple planet formation simulations to estimate the types and locations of planets that could exist in the Alpha Centauri system, similar to those in \cite{2002ApJ...576..982Q} etc. We begin from the planetesimal stage and simulate collisional accretion for 100 Myrs, until a stable population results. Extensive studies of planet formation in the modern Alpha Centauri system and other binaries have been done (e.g. see \cite{2007ApJ...660..807Q}), and our goal here is not to supersede or recreate them, but rather to test the strength of the influence of uncertainty in the initial disk model's physical parameters in comparison with small differences in truncation radius. Collisional modeling has also advanced significantly, e.g. \cite{2012ApJ...745...79L}, but such models are computationally expensive and introduce many new parameters, and as such are beyond the scope of this study, though would be appropriate for follow up studies. We focus in this section on planets around Alpha Centauri B as it has been the focus of more planet detection activity, but the above disk simulations show the environment around Alpha Centauri A to be qualitatively similar.

%- - - - - - - - - - - - - - - - - - - - - - - - - - - - - - - - - - - - - - - 
\begin{figure*}
\begin{center}
\includegraphics[scale=1.0]{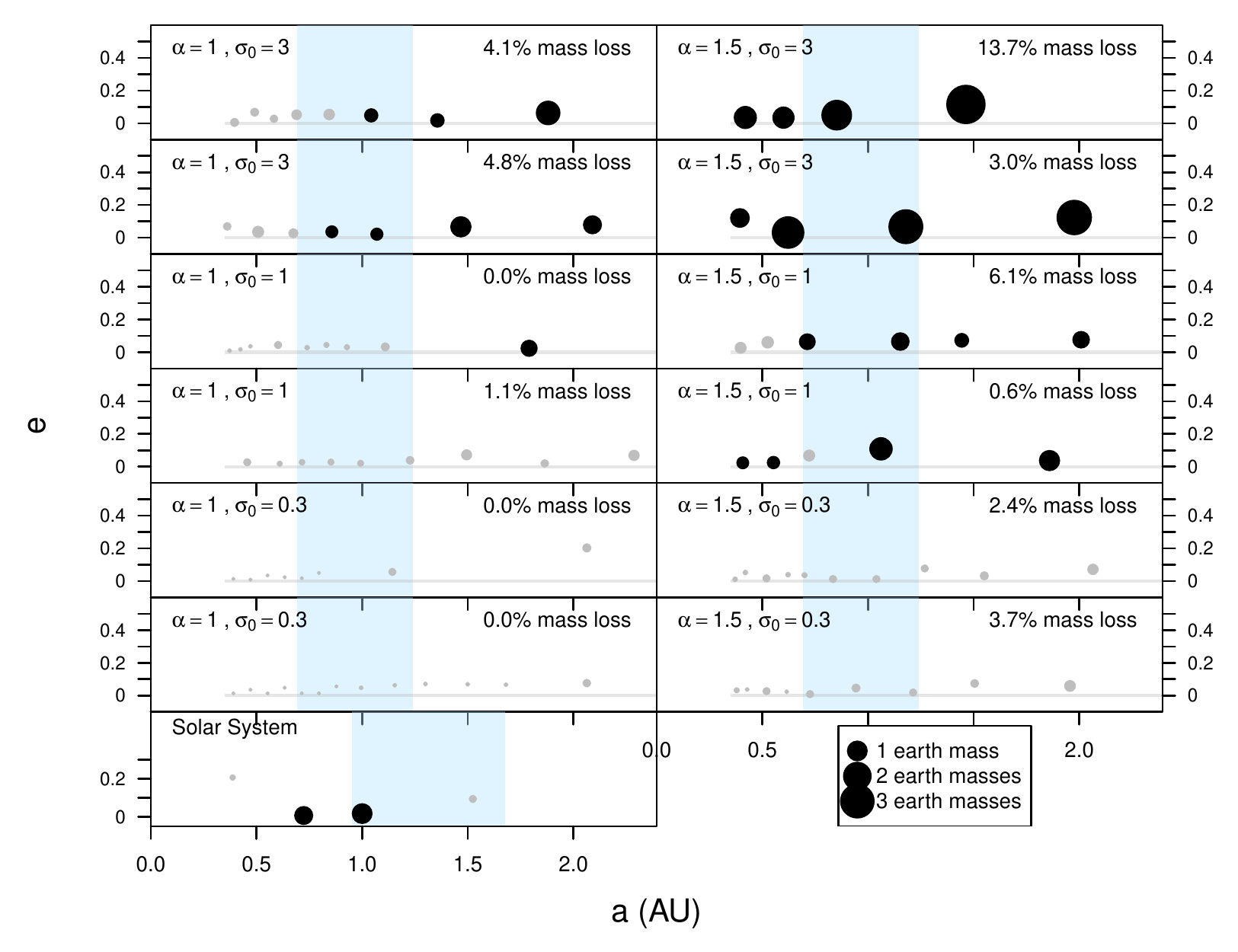}
% Created by PlanetSims/ReadElem.R
\caption{Semimajor axis and eccentricity of objects formed in the simulations with current Alpha Centauri binary orbital parameters and $r_{tr} = 2.5$ AU, representing the system after interactions with Proxima Centauri. The symbol size is proportional to the object's radius, and the bottom left panel shows the inner Solar System for comparison. Planets larger than 3M$_{Mars}$ (0.32 M$_{\oplus}$) are plotted in black, and smaller objects in gray. Blue shaded region indicates the traditional conservative habitable zone. Grey shaded bars along the bottom of each plot show the original extent of the disk. The percentage by mass of the initial disk ejected or accreted onto the stars during planet formation is given on the top right. Each plot is labeled with the disk parameters in the top left, where the disk density scales with $\sigma_0$, and $\alpha=1.5$ is the power law slope. $\alpha=1.5$ and $\sigma_0 = 1$ is the Minimum Mass Solar Nebula. \label{PltSystemsLow}}
\end{center}
\end{figure*}
%- - - - - - - - - - - - - - - - - - - - - - - - - - - - - - - - - - - - - - - 
\begin{figure*}
\begin{center}
\includegraphics[scale=1.0]{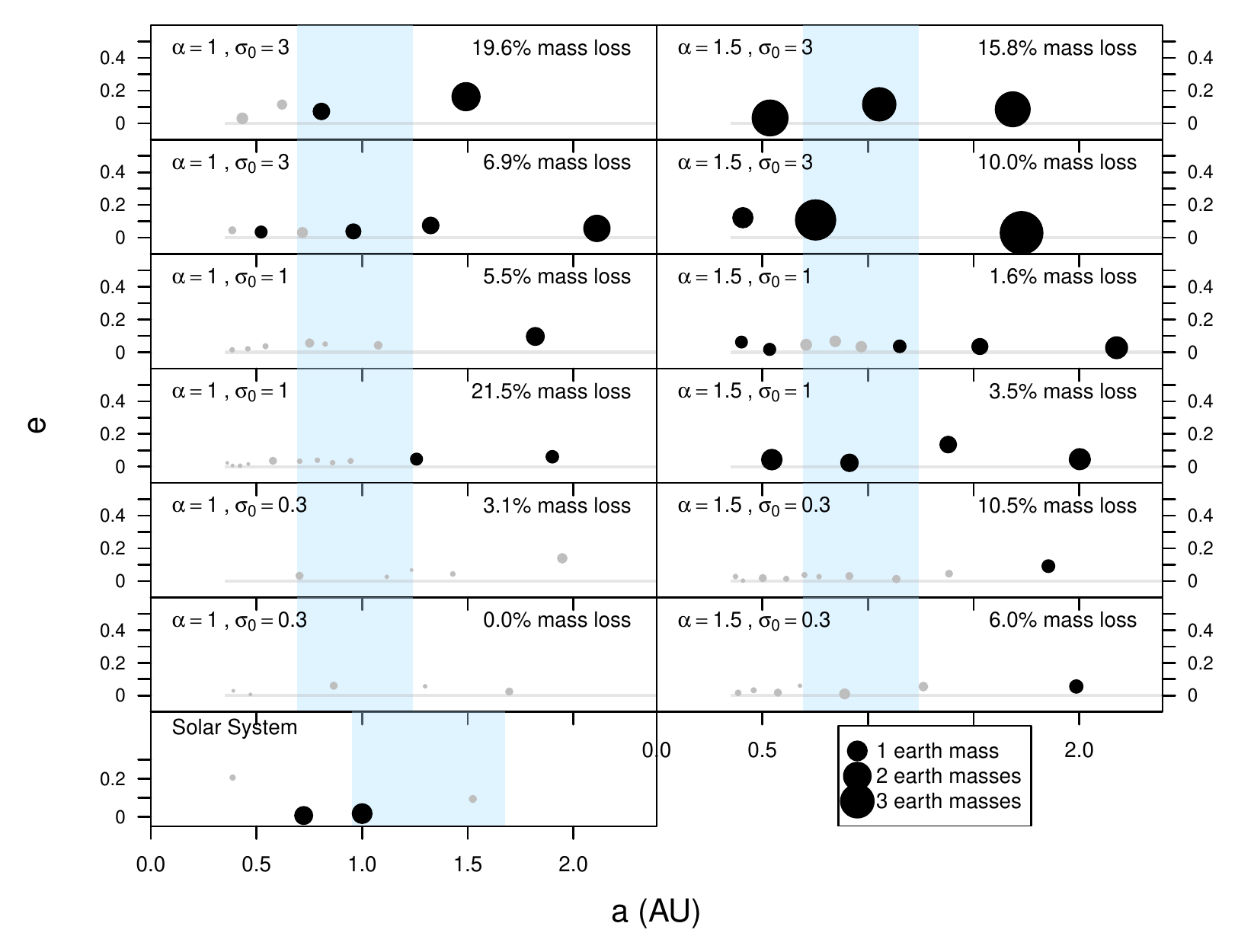}
% Created by PlanetSims/ReadElem.R
\caption{Same as Fig.~\ref{PltSystemsLow} but for $r_{tr} = 2.8$ AU, representing the system assuming no interaction with other stars. \label{PltSystemsMed}}
\end{center}
\end{figure*}
%- - - - - - - - - - - - - - - - - - - - - - - - - - - - - - - - - - - - - - - 
\begin{figure*}
\begin{center}
\includegraphics[scale=1.0]{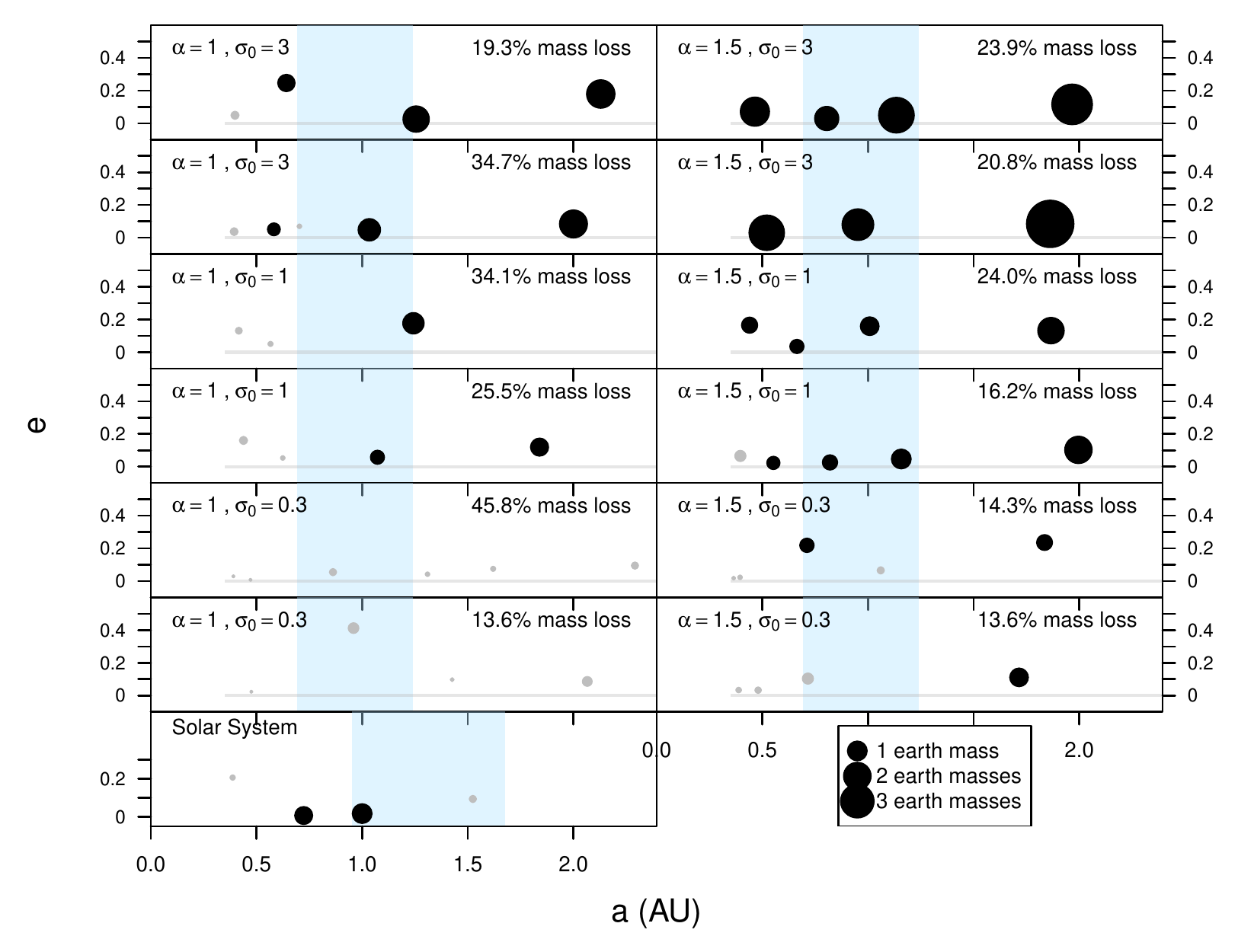}
% Created by PlanetSims/ReadElem.R
\caption{Same as Fig.~\ref{PltSystemsHigh} but for $r_{tr} = 3.1$ AU. The similarity to Fig.~\ref{PltSystemsMed} in planets formed and the higher ejection rates confirm that material placed beyond the truncation radius of 2.77 AU is quickly removed. \label{PltSystemsHigh}}
\end{center}
\end{figure*}
%- - - - - - - - - - - - - - - - - - - - - - - - - - - - - - - - - - - - - - - 
\begin{table}[htb]
\caption{Mean number of planets per simulation \label{NumPlts}}
\begin{center}

%%%% All plts
\begin{tabular}{l|ll|ll}

 &            \multicolumn{2}{c}{\textbf{All}} & \multicolumn{2}{|c}{\textbf{HZ}} \\
\backslashbox{\boldmath$\sigma_0$}{\boldmath$\alpha$} &    1 &  3/2 &    1 &  3/2 \\
\hline
% from PltsPerSystem.tex, made by PltSims/CountPlts.py
                3   &  3.2 & 3.5  &  1.5 & 0.8 \\
                1   &  1.2 & 4.2  &  0.5 & 1.3 \\
                0.3 &  0.0 & 0.8  &  0.0 & 0.0 \\
\end{tabular}
\tablecomments{The number per simulation of final objects with mass at least 3 M$_{Mars}$, averaged over simulations with each given set of parameters. Left two columns show the number per system, while the right two show number in the habitable zone per system.}
\end{center}
\end{table}

%- - - - - - - - - - - - - - - - - - - - - - - - - - - - - - - - - - - - - - - 

The initial configuration is a planetesimal disk around Alpha Centauri B with half the mass in Moon-sized objects and the other half in Mars-sized objects. All planetesimals initially have zero eccentricity and inclination, and are spaced such that the density of the disk follows Eqn.~\ref{SigProfile} out to some truncation radius $r_{tr}$. Alpha Centauri A orbits the system at its current distance and eccentricity. This configuration represents the system after it has interacted with Proxima Centauri -- the stars have attained their current orbital configuration and the truncated disk has evolved into planetesimals and planetary embryos. Proxima is by this point too distant to have a noticeable effect and is therefore omitted. 

We vary three parameters to compare their relative significance to planet formation: $\alpha$ takes the values 1 or 3/2; $\Sigma_0$ is multiplied by either 3, 1, or 0.3; and $r_{tr}$ values of 2.5, 2.8, and 3.1 AU were tested. The first two $r_{tr}$ values are those predicted for the Alpha Centauri system with and without Proxima interactions in Section 3.2, above. The set of simulations using the third value, $r_{tr} = 3.1$ AU, was created based on preliminary calculations, but is included because it further reinforces the truncation radius relationship found in the disk simulations above: a significantly higher fraction of the disk mass was ejected from disks containing material beyond the expected outer radius, leaving systems otherwise broadly similar to those of the $r_{tr} = 2.8$ AU set. For each combination of parameters, two simulations were run, all of which can be seen in Figs.~\ref{PltSystemsLow}--\ref{PltSystemsHigh}. 

We are using the arbitrarily-chosen mass cutoff of at least 3 M$_{Mars}$ as the threshold to consider objects planets, to separate out unprocessed disk material; the number of planets per simulation based on this metric can be seen in Table~\ref{NumPlts}. Simulations with at least the density of the MMSN resulted in systems of multiple terrestrial planets, frequently including planets in the habitable zone, while low-density simulations might form only one object large enough to be considered a planet by our definition. We defined the HZ boundaries as 0.693 to 1.241 AU, based on the conservative limits from Ravi Kopparapu's habitable zone calculator \citep{2013ApJ...765..131K, 2014ApJ...787L..29K} using Alpha Cen B's parameters. For a planet in this region, at closest approach, Alpha Cen A would contribute no more than 5\% the flux of B, which would move the bounds of the HZ inward by a small amount. Therefore, we use the single-star HZ as an approximation for this study; however, for further discussion of the binary's influence see \cite{2012MNRAS.422.1241F}.

The characteristics of the final system were most strongly dependent on the disk model parameters, particularly density. The small difference in truncation radius is unimportant in comparison. In systems with 0.3 times the MMSN, the debris remained in many small objects, and all of the final bodies were smaller than one M$_\oplus$. Systems with 1 or 3 MMSN formed several Earth or super-Earth sized planets, but in fewer numbers. Little disk material survives beyond the ice line, so the planets may be lower in volatiles and will likely resemble terrestrial planets rather than mini-Neptunes. However, the water fraction for a planet like the Earth is quite low compared to that of material beyond the ice line, so we do not consider this a barrier to habitability.

\subsection{Planets around Proxima}

Although we did not explicitly perform simulations of disks or planet formation around Proxima Centauri, we can make some predictions in this regime. Proxima's closest approach during its interaction with the binary ranges from several hundred AU to within the binary's orbit, though it tends to be on the larger end. If we assume that the trend in disk truncation relative to minimum pericenter continues linearly at larger separations with $r_{tr} \approx 0.3p_{min}$, then in the typical case where Proxima remained several hundred AU from Alpha Centauri, it could retain disk material within several tens of AU, sufficient to form planetary systems though possibly tending towards somewhat smaller planets than it would otherwise have. If, however, the star made a close pass of the inner binary, it likely would have lost most or all of its disk material. If future observations are able to rule out planetary companions to Proxima Centauri, this may be an indicator of past close interactions within the triple star system. Detection of planets could be used to put a lower limit on close approach, in degeneracy with outward migration.

%%%%%%%%%%%%%%%%%%%%%%%%%%%%%%%%%%%%%%%%%%%%%%%%%%%%%%%%%%%%%%%%%%%%%%%%%%%%%%%
\section{Conclusions}

In a large suite of N-body simulations, we recreate possible pathways to the formation of the Alpha Centauri-Proxima Centauri system according to the method described in \cite{2012Natur.492..221R}, and explore the effects this would have on planet formation around the individual stars. In typical scenarios, assuming a debris disk at least as dense as the Minimum Mass Solar Nebula, we expect several terrestrial planets to form within a few AU of the central star, with a high probability of a planet in the habitable zone.

We find it is plausible that Proxima formed at a distance of a few hundred AU from Alpha Centauri and was thrown out to the current (presumed) orbit at tens of thousands of AU while the stars were still accreting mass. In the process, it most likely caused the Alpha Centauri binary to see a decrease in semi major axis and increase in eccentricity, reducing the pericenter. This would cause truncation of the outer protoplanetary disk to a radius that scales with the binary's pericenter. Binaries which have undergone this type of three-body interaction in the past may have gone through a phase with a tighter orbit than it is now; the minimum pericenter could take any value up to the present one, but is typically no more than 20\% smaller). This additional truncation strips additional outer material from the disk, but in most simulations, enough material remains to form systems of terrestrial planets. Although the process of planetesimal formation, especially in binaries, is not yet fully understood, we make the assumption that the system was able to form planetesimals and planetary embryos, and then examine how planet formation would proceed from that point.

In our simulations of the Alpha Centauri system, the disk was typically truncated to near the ice line, likely preventing formation of gas giants and even Neptunes or mini-Neptunes. See \cite{2013MNRAS.431.3444C}, \cite{2014MNRAS.440L..11R}, and \cite{2014ApJ...780...53C} for discussions of in situ versus migration formation pathways of close-in mini-Neptunes. Unless the initial disk had a much higher density than expected, high mass, close-in planets such as some found by Kepler \citep{2014ApJ...790..146F} are not expected here, with neither an outer region to form planets that migrate in, nor an outer well of material to feed an inner region while planets form in situ.

In addition, these interactions do not necessarily rule out the possibility of planets around Proxima, although an absence of planets may indicate a past close encounter and serve as a confirmation of triple system interactions involving a close pass.

Finally, we predict that if Proxima is orbiting at high inclinations, it may be inducing Kozai-Lidov oscillations in Alpha Centauri which will change its eccentricity on a Gyr timescale. It is uncertain whether this would increase or decrease the pericenter from the present value, but it most likely will not decrease it further than the minimum pericenter it has experienced previously, and so should not cause significant disturbance beyond the previous truncation. If, however, the outer star is perturbed enough by external forces such as passing stars, it could end up on a new orbit, essentially randomizing the system.

Overall, significant uncertainties remain, but our simulations indicate that, despite the possibility of a turbulent past, Alpha Centauri B and its companions are still likely terrestrial planet hosts. Missions capable of detecting or conclusively ruling out such planets would yield great insights into the formation of planets within multistellar environments. Any planets found would provide targets for detailed characterization, while a non-detection would be a good indicator of the system having undergone disruptive stellar interactions, which helps constrain the fitness of multistellar systems as planet hosts. Therefore, we look forward to results from current and future searches of this system.

%%%%%%%%%%%%%%%%%%%%%%%%%%%%%%%%%%%%%%%%%%%%%%%%%%%%%%%%%%%%%%%%%%%%%%%%%%%%%%%
\section{Acknowledgments}
The authors thank the The Pennsylvania State University, NASA Astrobiology Institute (NNA09DA76A), and the Penn State Astrobiology Research Center for their support. Portions of this research were conducted with Advanced CyberInfrastructure computational resources provided by The Institute for CyberScience at The Pennsylvania State University (http://ics.psu.edu). Special thanks go to Prof. Eric Ford for directing us to Dimitri Veras' work on including Galactic tides in {\sc mercury}. In addition, we thank the anonymous referee for thorough and helpful comments which contributed to the content and clarity of this work.

%%%%%%%%%%%%%%%%%%%%%%%%%%%%%%%%%%%%%%%%%%%%%%%%%%%%%%%%%%%%%%%%%%%%%%%%%%%%%%%

\bibliography{AlphaCen_ADS}

%%%%%%%%%%%%%%%%%%%%%%%%%%%%%%%%%%%%%%%%%%%%%%%%%%%%%%%%%%%%%%%%%%%%%%%%%%%%%%%
\clearpage

%%%%%%%%%%%%%%%%%%%%%%%%%%%%%%%%%%%%%%%%%%%%%%%%%%%%%%%%%%%%%%%%%%%%%%%%%%%%%%%
\end{document}